\documentclass[preprintnumbers,floatfix,superscriptaddress,nofootinbib]{revtex4}
\usepackage{amsbsy}
\usepackage{amsfonts}
\usepackage{amssymb}
\usepackage{amsmath}
\usepackage{graphicx,color}
\usepackage{graphicx}
\usepackage{subfigure}
\usepackage{epsfig}
\usepackage{graphicx} 
\usepackage{dcolumn} 
\usepackage{bm} 
\usepackage{color}
\usepackage{epstopdf}
\newcommand{\be}{\begin{equation}}
\newcommand{\ee}{\end{equation}}

\begin{document}

\newcommand{\ket}[1]{{\left\vert {#1} \right\rangle}}
\newcommand{\bra}[1]{{\left\langle {#1} \right\vert}}

\title{Quantum coherent contributions in biological electron transfer}
\author{Ross Dorner}
\affiliation{Controlled Quantum Dynamics Theory, Department of Physics, Imperial College London, SW7 2AZ, U.K}
\affiliation{Atomic and Laser Physics, Clarendon Laboratory, University of Oxford, Parks Road, University of Oxford, OX1 3PU, U.K.}
\author{John Goold}
\affiliation{Atomic and Laser Physics, Clarendon Laboratory, University of Oxford, Parks Road, University of Oxford, OX1 3PU, U.K.}
\affiliation{Department of Physics, University College Cork, Cork, Ireland}
\author{Libby Heaney}
\affiliation{Atomic and Laser Physics, Clarendon Laboratory, University of Oxford, Parks Road, University of Oxford, OX1 3PU, U.K.}
\affiliation{Centre for Quantum Technologies, National University of Singapore, 3 Science Drive 2, 117543, Singapore}
\author{Tristan Farrow}
\affiliation{Centre for Quantum Technologies, National University of Singapore, 3 Science Drive 2, 117543, Singapore}
\author{Phillipa G. Roberts}
\affiliation{Medical Research Council Mitochondrial Biology Unit, Wellcome Trust/MRC Building, Hills Road, Cambridge, CB2 0XY, U.K.}
\author{Judy Hirst}
\affiliation{Medical Research Council Mitochondrial Biology Unit, Wellcome Trust/MRC Building, Hills Road, Cambridge, CB2 0XY, U.K.}
\author{Vlatko Vedral}
\affiliation{Atomic and Laser Physics, Clarendon Laboratory, University of Oxford, Parks Road, University of Oxford, OX1 3PU, U.K.}
\affiliation{Centre for Quantum Technologies, National University of Singapore, 3 Science Drive 2, 117543, Singapore}
\maketitle
\section*{ABSTRACT}
Many biological electron transfer (ET) reactions are mediated by metal centres in proteins. NADH:ubiquinone oxidoreductase (complex I) contains an intramolecular chain of seven iron-sulphur (FeS) clusters~\cite{Sazanov:06}, one of the longest chains of metal centres in biology and a test case for physical models of intramolecular ET. In biology, intramolecular ET is commonly described as a diffusive hopping process, according to the semi-classical theories of Marcus and Hopfield~\cite{Marcus:85, Hopfield:74}.
However, recent studies have raised the possibility that non-trivial quantum mechanical effects play a functioning role in certain biomolecular processes~\cite{Engel:07, Gauger:11}.  
Here, we extend the semi-classical model for biological ET to incorporate both semi-classical and coherent quantum phenomena using a quantum master equation based on the Holstein Hamiltonian~\cite{Holstein:59}.  We test our model on the structurally-defined chain of FeS clusters in complex I.  By exploring a wide range of realistic parameters we find that, when the energy profile for ET along the chain is relatively flat, just a small coherent contribution can provide a robust and significant increase in ET rate (above the semi-classical diffusive-hopping rate), even at physiologically-relevant temperatures.  Conversely, when the on-site energies vary significantly along the chain the coherent contribution is negligible.  For complex I, a crucial respiratory enzyme that is linked to many neuromuscular and degenerative diseases~\cite{Lin:2006}, our results suggest a new contribution towards ensuring that intramolecular ET does not limit the rate of catalysis.  For the emerging field of quantum biology, our model is intended as a basis for elucidating the general role of coherent ET in biological ET reactions.




\section*{}
\begin{figure}[tb]
\includegraphics[width=0.5\linewidth]{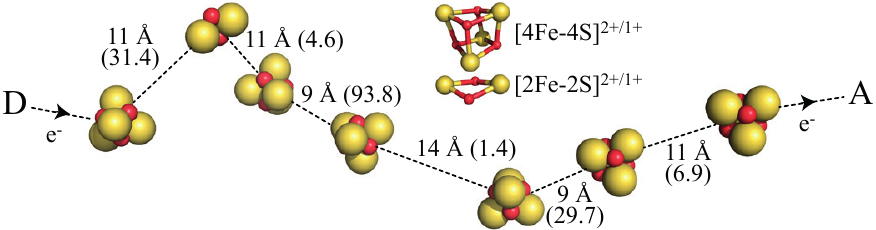}
\caption{The FeS clusters in complex I connect the NADH/flavin electron donor module (D) to the quinone electron acceptor (the electron sink, A).  Edge-to-edge distances between the clusters are marked in $\AA$, with tunneling amplitudes given in brackets in GHz. The structures of the 2Fe and 4Fe cores are inset.  The whole assembly is embedded in the protein matrix. Based on 2FUG.pdb~\cite{Sazanov:06}.}
\label{fig:rci}
\end{figure}

In NADH:ubiquinone oxidoreductase (complex I), the electron transfer (ET) reaction between NADH and quinone is mediated by seven iron-sulphur (FeS) clusters that link a flavin mononucleotide to the quinone binding site (see Fig.~\ref{fig:rci}).  Following reduction of the flavin by hydride transfer from NADH, electrons are transferred sequentially along the cluster chain, to the quinone.  The free energy released by the redox reaction is conserved in transmembrane proton translocation, in a separate domain of the enzyme.  Thus, ET along the cluster chain must be both fast-enough that it does not limit the rate of catalysis, and energetically efficient \cite{Hirst:11}.  
Here develop a general treatment of ET in metalloproteins, exploiting complex I as a template to elucidate whether quantum mechanical contributions play a role biological ET.
In our model, each cluster in the chain can exist in two states, oxidised or (one-electron) reduced, and each electronic state has a set of vibrational states associated with it.  We preserve essential features of the Marcus and Hopfield theories \cite{Marcus:85, Hopfield:74}, in which vibrations and vibronic couplings play a central role, by basing our model on the closely related Holstein Hamiltonian \cite{Holstein:59}
\begin{equation}
H= \sum_{i}^N E_i c_i^\dagger c_i+h \sum_{j}^{N-1} t_{j,j+1} \left( c^\dagger_j c_{j+1} +c^\dagger_{j+1} c_j \right)+ h \sum_{i}^N\nu_{i} a^\dagger_{i} a_{i}+h \sum_{i}^N g_{i} c^\dagger_{i} c_{i} \left(a_{i} + a^\dagger_{i} \right).
\label{eq:initham}
\end{equation}
In the first term of Eq.~\eqref{eq:initham}, $E_i$ is the on-site energy of an electron on the $i^{th}$ cluster ($E_i$ does not include interactions between the electron and cluster, see below), and $c_i^\dagger$ and $c_i$ are the electron creation and annihilation operators. The second term describes the coherent tunneling of the electron between the $j^{th}$ and $(j+1)^{th}$ clusters with amplitude 
$t_{j,j+1}$ ($h$ is Planck's constant). Coherent tunneling gives rise to a continuous wavelike motion of the electron, allowing it to exist in a quantum superposition of states across adjacent redox sites. This electron delocalisation is absent from diffusive models of ET, such as those of Marcus and Hopfield, in which an electron hops between well-defined locations with a certain probability.
The third term denotes the vibrational contribution to the energy of the system, where $\nu_{i}$ is the 
characteristic frequency of vibration of the $i^{th}$  cluster and the bosonic operators $a^\dagger_{i}$ and $a_{i}$ act to raise and lower, respectively,
the number of excitations in that vibrational mode. The Hamiltonian in Eq.~\eqref{eq:initham} describes the case of a single vibrational mode per site whereas, in reality, each FeS cluster has a large complement of 
modes. Generalisation of our model to a greater number of modes is possible using the approach outlined in the appendix.

Associated with the Holstein Hamiltonian is the concept of the polaron, a bosonic quasiparticle
formed by the electron and the local vibrational modes of the FeS cluster. Polaron formation arises from vibronic coupling, through which the reduction of a given cluster leads to a change in its vibrational state; it is described by 
the fourth term of Eq.~\eqref{eq:initham} for a vibronic coupling strength of $g_{i}$ at the $i^{th}$ cluster. The polaron concept simplifies
both the physical and mathematical picture considerably by treating the charge and its accompanying vibrational modes as a single entity. The concept is 
made mathematically explicit by the Lang-Firsov transformation~\cite{Lang:63} (see appendix), allowing Eq~\eqref{eq:initham} to be re-expressed, with renormalised parameters, by Eq.~\eqref{eq:effham}.
\begin{equation}
\bar{H}= \sum_i^N E^*_i c^\dagger_i c_i+h\sum_{j}^{N-1}t_{j,j+1} \left(c^\dagger_j c_{j+1} \bar{X}_{j}^\dagger \bar{X}_{j+1}
+c^\dagger_{j+1} c_j \bar{X}_{j+1}^\dagger\bar{X}_{j}\right)
+ h\sum_i^N \nu_{i} a_{i}^\dagger a_{i}.
\label{eq:effham}
\end{equation}

Where $E_i^*=E_i- \Delta_i$ is the reduction potential of the $i^{th}$ cluster, $\bar{X}_{i}= \textrm{exp}\left(-{\frac{g_{i}}{\nu_{i}}\left(a_{i}^\dagger-a_{i} \right)}\right)$
and $\Delta_i =\frac{h g_{i}^2}{\nu_{i}}$.
Note that the Hamiltonian in Eq.~\eqref{eq:effham} is general enough to be applied to a wide range of metalloproteins in which ET depends strongly on a set of parameters that can only be extracted from experimental studies. In this study, complex I is used as a template to investigate the possibility of quantum mechanical contributions to biological ET reactions. Two reduction potential profiles were examined. First, when the seven-cluster chain in complex I contains four electrons they occupy alternate positions along the chain~\cite{Roessler:10}, suggesting an alternating reduction potential profile, with significant variation in potential between adjacent sites. Alternatively, the same pattern of occupancy could result from electrostatic interactions between the sites, suggesting a flatter underlying profile.  The relative contributions of these two effects remains unclear. For the alternating profile \cite{Hirst:11}, the $E_i^*$ values fluctuate significantly (on the order of 0.1 eV) from site-to-site.  For such large energetic disorder, the quantum mechanical contribution to ET is negligible and the dynamics are well described by purely diffusive modelling.  However, using a flat profile in which $E_i^*$ is taken to be equal for all clusters other than site 7 (that is known to have the highest reduction potential), the quantum mechanical contributions to ET become significant.  Thus, we have chosen this flat profile (with sites to 1 to 6 at 0 eV and site 7 at $-0.15$ eV (36.4 THz)) as the main focus of our study; the flat profile has also been adopted in other modelling studies~\cite{Hayashi:10, Moser:06}.

The vibrational modes of protein-bound FeS clusters have been characterised using nuclear resonance vibrational spectroscopy, resonance Raman spectroscopy and density functional theory (DFT)~\cite{Mitra:11}; we select a common 
frequency  $\nu_i=\nu=334$ $\textrm{cm}^{-1}$ (10 THz), $ i =1, 2 \hdots 7$ for all sites in the chain.
We estimate the vibronic coupling strength from DFT simulations of the so-called `inner sphere reorganisation energy' $\lambda_\textrm{in}$ 
by noting the equivalence of this quantity to $\Delta_i$. Ref.~\cite{Mitra:11} provides an inner sphere reorganisation 
energy for FeS clusters $\lambda_\textrm{in}=0.2$ eV (48 THz) to yield a common vibronic coupling strength for all sites, $g_i=g=\sqrt{\nu \lambda_\textrm{in}}=22 \textrm{ THz}.$


The tunnelling amplitudes are obtained from previous studies of the 
transfer integrals between FeS clusters within complex I using DFT and semi-empirical electronic structure methods~\cite{Hayashi:10}. Due to the disorder in 
spacing between clusters, amongst many other factors, these tunneling amplitudes vary between adjacent sites, introducing disorder to the Hamiltonian. The 
tunneling amplitudes all lie within the range $t_{j,j+1}=$ 1 - 95 GHz  (see Fig.~\ref{fig:rci} details), allowing the polaron tunnelling 
term of Eq.~\eqref{eq:effham} to be readily treated as a perturbation.

The presence of the bosonic operators $\bar{X}_i$ in Eq.~\eqref{eq:effham} gives rise to two distinct processes that contribute to the perturbation expansion~\cite{Holstein:59}. The first of these are so-called `diagonal transitions' in which the vibrational quantum numbers in the chain do not change upon the 
tunneling of an electron. Diagonal transitions facilitate quantum coherent ET and lead to a modulation of the tunneling amplitude between adjacent sites due to the effect of vibronic coupling
\begin{equation}
t^*_{j,j+1}=t_{j,j+1} \textrm{exp} \left(- \frac{g^2}{\nu^2} \left(1+2\langle n_\nu(T) \rangle\right) \right).
\label{eq:renhop}
\end{equation}
Where $\langle n_\nu(T)\rangle=\left(\textrm{exp}(h\nu/k_B T)-1\right)^{-1}$ is the thermal average of the number excitations in a vibrational mode 
with frequency $\nu$ at temperature $T$ and $k_B$ is Boltzmann's constant.
We thus define an effective Hamiltonian, correct to the first order (and assuming a single vibrational mode per cluster), 
describing the coherent transport of a polaron among the FeS clusters
\begin{equation}
H_\textrm{eff} =\sum_{i}^N E^*_i c_i^\dagger c_i+h\sum_{j}^{N-1} t^*_{j,j+1} \left( c^\dagger_j c_{j+1} +c^\dagger_{j+1} c_j \right)+h \nu\sum_{i}^N  a^\dagger_{i} a_{i}.
\label{eq:transham}
\end{equation}
The effect of the vibronic coupling has been wholly absorbed into the parameters $t_{j,j+1} ^*$ and $E_i^*$, resulting in exponential suppression of 
the tunnelling amplitude and a modulation of each on-site energy due to the deformation of the cluster by an electron. The second contribution 
to the perturbation series arises from `non-diagonal transitions' in which the vibrational quantum number may change. The quantum 
coherence of an electron is not preserved in non-diagonal transitions, allowing 
them to be described stochastically by a set of hopping rates that lead to diffusive motion. These rates are calculated within our model using the Marcus-Jortner equation~\cite{Jortner:76}
\begin{equation}
k_{j \rightarrow j+1}=2 \pi h \vert t_{j,j+1} \vert^2 \sqrt{\frac{\pi}{\lambda_\textrm{in}k_B T}} \textrm{exp}\left(-\frac{\lambda_\textrm{out}}{h \nu_\textrm{o}} \right)\sum_{m=0}^\infty \frac{(\frac{\lambda_\textrm{out}}{h \nu_\textrm{o}})^m}{m!} \textrm{exp}\left(\frac{-(E^*_j-E^*_{j+1}+mh\nu_\textrm{o} +\lambda_\textrm{in})^2}{4\lambda_\textrm{in}k_B T}\right),
\label{eq:marcjort} 
\end{equation}
and by enforcing detailed balance for the reverse rates. Here, $\nu_\textrm{o}=$10 THz is a typical frequency for the optical vibrational modes of the environment and $\lambda_\textrm{out}$ is the outer sphere reorganisation energy describing the interaction of the polaron with its surrounding protein scaffold and solvent. 
There is no consensus value for $\lambda_\textrm{out}$ for FeS clusters in the existing literature. We assume that the outer and 
inner sphere reorganisation energies are of comparable importance by choosing $\lambda_\textrm{out}=\lambda_\textrm{in}=0.2$ eV  (48 THz).

The incoherent hopping rates  Eq.~\eqref{eq:marcjort} enter into the description of  ET within the chain  as Liouvillian terms within a Born-Markov quantum master equation alongside the Hamiltonian dynamics (Eq. \eqref{eq:transham}) (see appendix for details).
Further terms are included in the master equation to address the fact that the closed system we have considered thus 
far is an inadequate description of `in situ' ET. In complex I, the FeS clusters are embedded within the
 protein scaffold and surrounding solution that constitutes their environment. Interaction of a quantum system with its environment introduces non-reversible processes to its dynamics; these 
manifest themselves as local dephasing - a process which reduces quantum superposition states to classical statistical mixtures by the process of
`decoherence' at a given rate. In addition, we include the irreversible dissipation of the polaron from the final FeS cluster of the chain to the
quinone at a rate $R=1$ GHz that is of the order of other ET processes in the chain.
The origin of local dephasing can be associated with the outer sphere 
reorganisation energy $\lambda_\textrm{out}$. In lieu of a detailed knowledge of the density of modes 
for the environment, we assume a dephasing rate of the form
\begin{equation}
\gamma=\gamma_0(1+\langle n_{\nu_\textrm{a}}(T)\rangle ) \approx \gamma_0\left(1+\frac{k_B T}{h \nu_\textrm{a}}\right),
\label{eq:dephrates}
\end{equation}
where $\nu_\textrm{a}=0.1$ THz is taken as a typical frequency for the vibrational modes of the environment that are responsible for dephasing~\cite{Mckenzie:08}
and $\gamma_0$ is the projected magnitude of quantum dephasing at zero temperature.
Considering Eqs.~\eqref{eq:renhop},~\eqref{eq:marcjort} and~\eqref{eq:dephrates}, we see that with increasing temperature the amplitude
of coherent tunneling is reduced as the rate of incoherent tunneling and local dephasing increases.

The time evolution of the system is obtained by standard numerical integration of the Born-Markov master equation with the boundary condition that the electron 
is initially localized at the first site of the chain. The motion of the electron among the seven site chain is governed by a number of factors, notably; 
the disorder of the chain - which here enters via the tunneling amplitudes - and the temperature of the system. At low temperature, the coherent (diagonal) contribution to ET is significant compared to 
the rate of incoherent (non-diagonal) processes. As the temperature is increased, the importance of coherent transport is reduced due to the suppression of the coherent tunneling amplitudes and the onset of dephasing and incoherent tunneling, leading to diffusion-like transport. For high enough values of the zero temperature dephasing ($\gamma_0 \sim 1$ GHz) at $T=300$ K, quantum coherence is no longer maintained on a timescale relevant to ET and purely diffusive transport is recovered with rates given by the Marcus-Jortner equation (see appendix).
\begin{figure}[tb]
\includegraphics[width=0.8\linewidth]{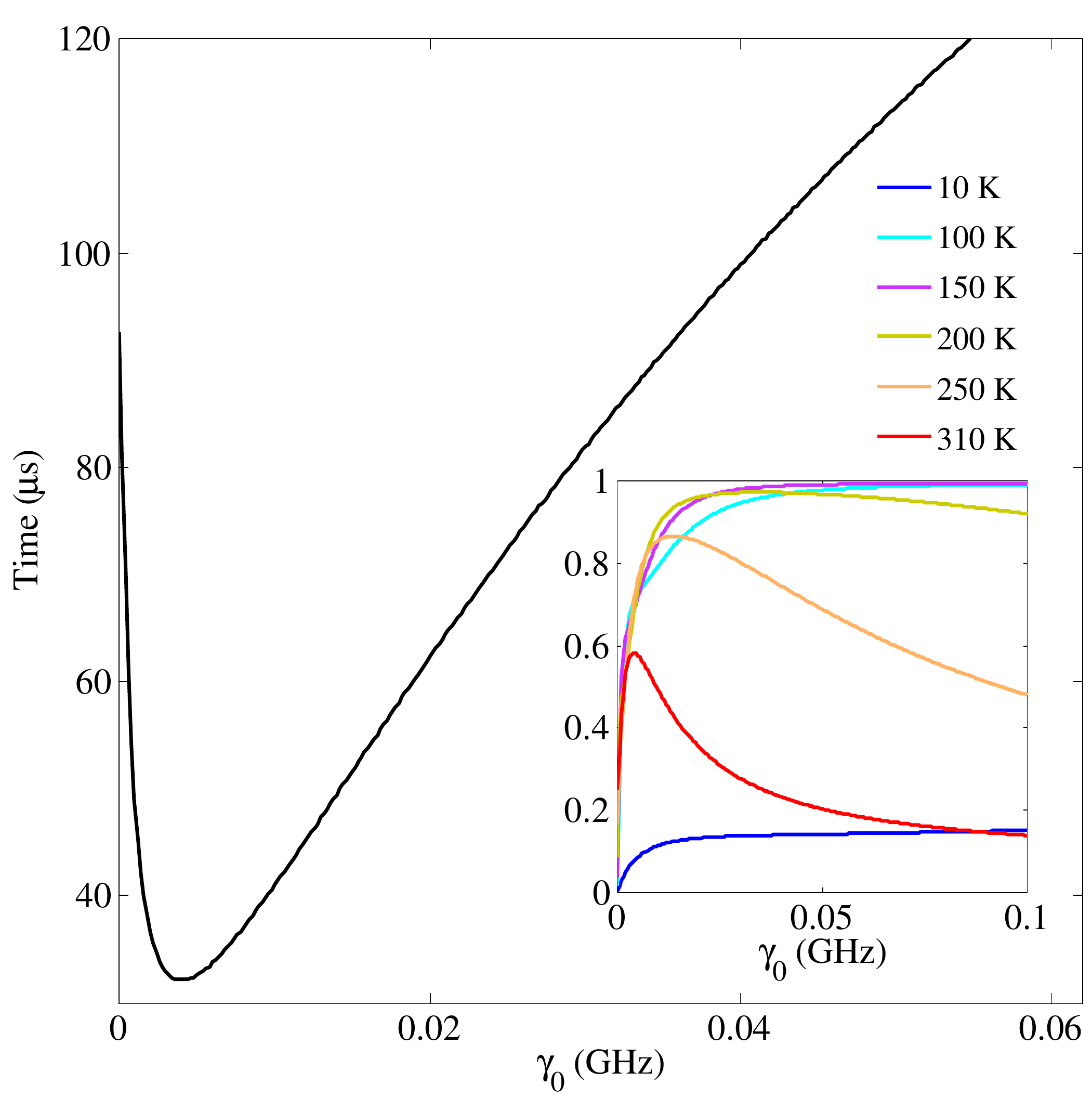}
\caption{The time taken to achieve 50 $\%$ probability of quinone reduction (50$\%$ population of the sink) at 310 K, after the electron was initialised in site 1, as a function of zero temperature dephasing $\gamma_0$. The shortest time for quinone reduction, 35 $\mu$s, is observed at $\gamma_0 =$ 0.004 GHz. This is almost 10 times shorter than the analogous semi-classical treatment using the
Marcus-Jortner equation (304 $\mu$s). For comparison, diffusive modeling using the Marcus equation \cite{Marcus:85} predicts a time of 136 $\mu$s
and the empirical treatment of Dutton et al.~\cite{Moser:06} estimates the time required to be 7 $\mu$s (for a reorganisation energy of 0.4 eV). {\bf Inset:} The probability of quinone reduction  is shown at $t=40$ $\mu$s 
as a function of $\gamma_0$ for various temperatures.}
\label{fig:fig1}
\end{figure}
Fig.~\ref{fig:fig1} shows the sensitivity of the transport to the temperature and amount of zero temperature dephasing $\gamma_0$.

At 10 K there is little probability for the electron to reduce the quinone with any level of dephasing. 
This is a feature of the reduction potential profile of the chain:
While sites 1-6 have similar on-site energies, the energy of site 7 is $0.15$ eV ($36.4$ THz) lower. 
This difference in energy is much greater than the tunneling amplitude between sites $6$ and $7$. 
Consequently, coherent transport is strongly suppressed between these sites and occupation of site $7$ is achieved 
by non-diagonal terms that become more pronounced at higher temperatures.
Using this profile, the final step in the ET chain of complex I is thus a thermally activated incoherent hop, though transport elsewhere 
in the chain may be quantum coherent. The alternating profile discussed above yields purely incoherent dynamics for similar reasons.

\begin{figure}[tb]
\includegraphics[width=0.82\linewidth]{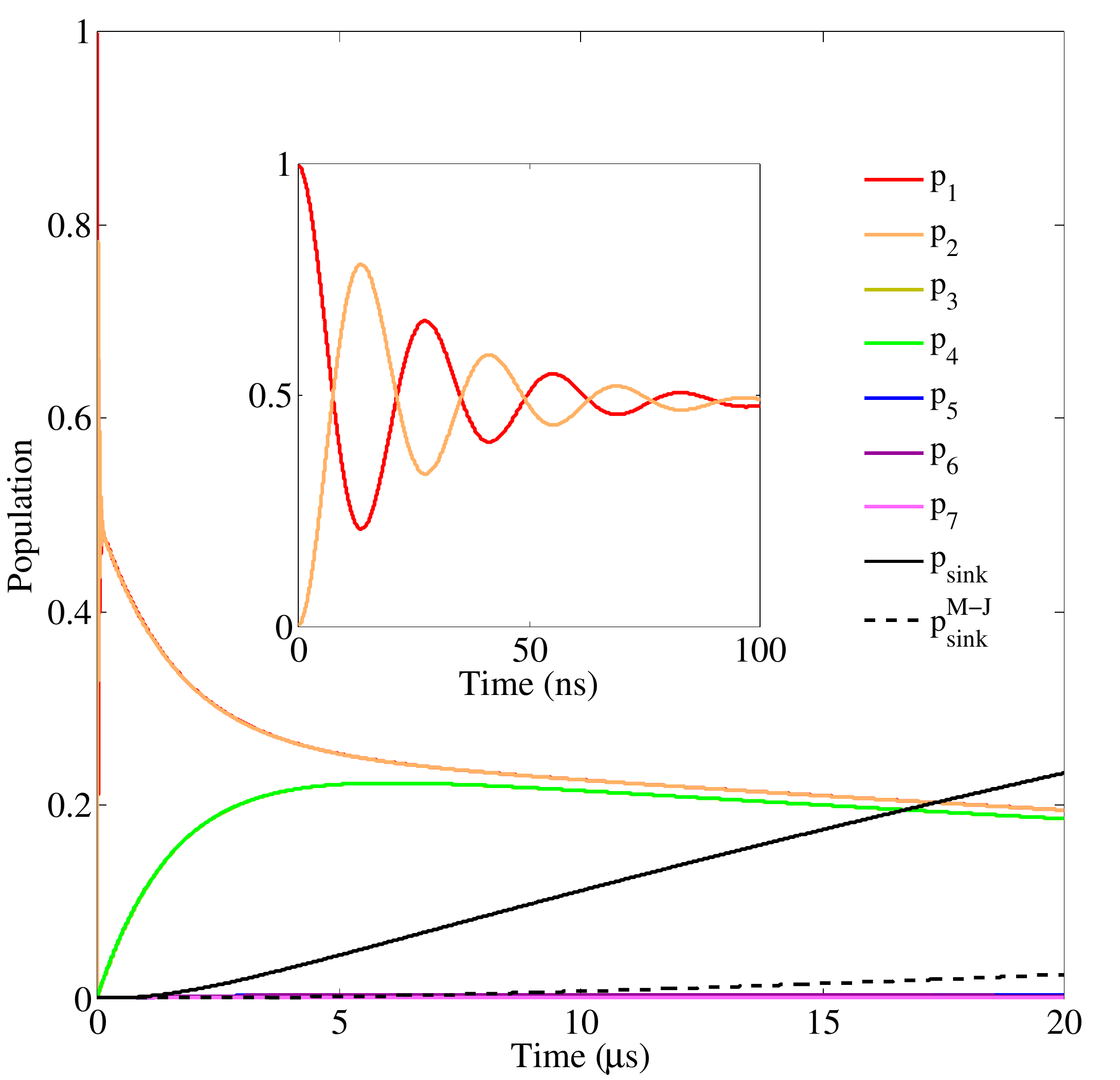}
\caption{The populations  of the seven sites of the FeS cluster chain in complex I at $T=310$ K and $\gamma_0=0.001$ GHz. 
The timescale for ET to the sink (reduction of the quinone) is roughly 10-100 $\mu$s, over which the ET appears diffusive. The populations  of sites 1~\&~2, and  3~\&~4, are overlaid, while sites 5 - 7 are barely ever populated.
Much shorter timescales (100 ns) show the presence of quantum coherent transport between the first two sites of the chain (inset). For comparison, the population of the sink is shown for the case of purely diffusive dynamics calculated using the Marcus-Jortner equation (dashed line).}
\label{fig:fig2}
\end{figure}

\begin{figure}[tb]
\centering
\includegraphics[width=0.8\linewidth]{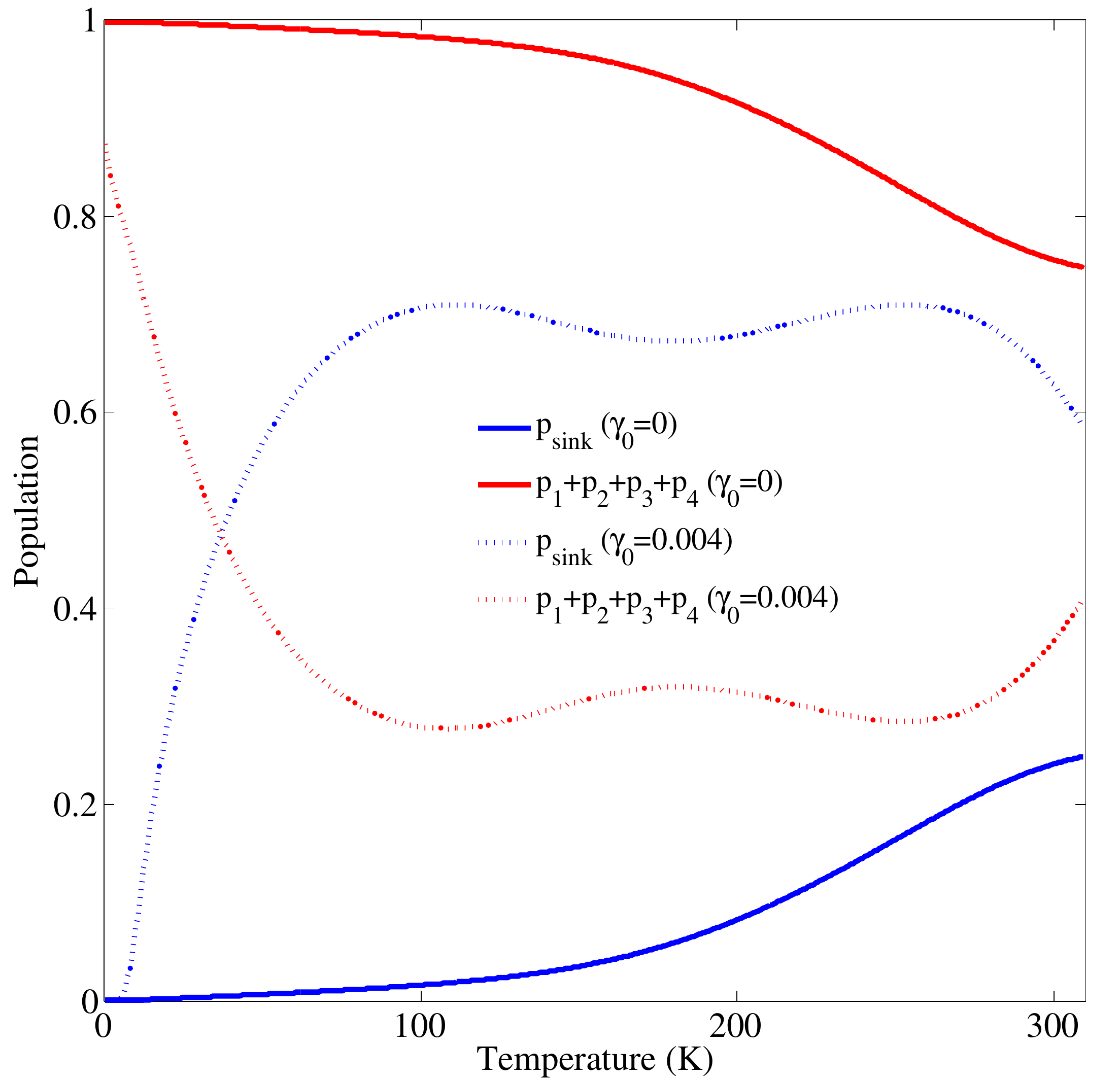}
\caption{The population of the first four sites of the cluster chain and the quinone as a function of temperature for $\gamma_0=0$ (solid lines) and `optimal' dephasing ($\gamma_0=0.004$ GHz, dotted lines) after 40 $\mu$s.}
\label{fig:fig3}
\end{figure}

The rate of ET from site 1 of the chain to the quinone molecule increases between 10 - 150 K 
across the full range of $\gamma_0$ that was investigated. Above 200 K the transport becomes increasingly sensitive to the value of $\gamma_0$. At 310 K the fastest 
electron transfer to the quinone is achieved for $\gamma_0 \approx 0.004$ GHz (see Figs.~\ref{fig:fig1}~-~\ref{fig:fig3}) but for other values of $\gamma_0$ the transport is far less efficient. These features are akin to the `dephasing assisted transport'~\cite{Giorda:11, Chin:2010, Rebentrost:09} 
that has been much discussed in the context of quantum coherent exciton transport in photosynthetic complexes.
Dephasing assisted transport predicts that an intermediate region between purely coherent and incoherent dynamics offers the fastest rate of transport. 
Notably, the behaviour may 
not be solely attributed to dephasing, due to the presence of incoherent tunneling, but the notion of ET enhancement by the interplay of quantum coherent and incoherent processes remains.

The presence of dephasing assisted transport, and hence partly quantum coherent ET within  complex I, at physiological 
temperature and in certain regimes of the parameter $\gamma_0$, is further explored in Fig.~\ref{fig:fig2}, 
which shows the population of each of the 7 sites of complex I as a function of time for $\gamma_0 \approx 0.001$ GHz at 310 K (similar plots for different values of $\gamma_0$ are given in the appendix).
On the timescale of ET from site 1 to the quinone molecule (10 - 100 $\mu$s) the transfer appears qualitatively diffusive. 
The electron quickly equilibrates among the first four sites - which are to a certain extent decoupled from the rest of 
the chain due to the large distance between sites 4 and 5 - by thermally activated hopping. Upon making a transition from site 4 to site 5, the electron 
is more likely to proceed toward the sink than to undergo a transition back to site 4. The disorder in the tunneling amplitudes therefore introduces 
an effective directionality to the electron motion. Further directionality arises from the thermally activated hop between sites 6 and 7, and the 
low population in sites 5 - 7 may be described by the electron quickly being transferred to the sink upon transition from site 4 to 5. 
On much shorter timescales (100 ns), coherent oscillations of the electron persists between sites 1 and 2 of the chain (see inset to Fig.~\ref{fig:fig2}) 
and to a lesser extent between sites 3 and 4. These oscillations are the hallmark of quantum coherence and are contrary to the typical relaxation-like behaviour 
exhibited in classical transport. Within our model, coherent oscillations can persist for upwards of 100 ns for $\gamma_0\leq 0.004$ GHz. 
Remarkably, despite quantum coherence only persisting on a timescale two orders of magnitude slower than that of the overall ET, quantum effects have a significant effect on the transport rate, yielding roughly ten times the probability of quinone reduction after 40 $\mu$s compared to the analogous purely diffusive Marcus-Jortner model.
To further emphasise this feature, Fig.~\ref{fig:fig3} shows the population of the 
quinone molecule and sites 1 - 4 (as sites 5 - 7 are only ever sparsely populated) when the chain is subject to different rates of dephasing. For no dephasing ($\gamma_0=0$), 
the probability of quinone reduction does not exceed 0.25 within 40 $\mu$s for $T\leq310$ K. However, for $\gamma_0 = 0.004$ GHz
the rate of transfer is increased at all temperatures, with a maximum probability of roughly 0.7 for quinone reduction after 
40 $\mu$s. This is evidence that the addition of local dephasing processes can enhance
ET that is already partly incoherent due to non-diagonal processes.

We have presented a full microscopic quantum model of ET in metalloproteins using a novel approach based on the Holstein Hamiltonian~\cite{Holstein:59} and a quantum master equation. By accounting for the unique properties of quantum dynamics up to the first order of perturbation theory, we show that, in principle, even small quantum mechanical contributions can enhance ET rates in metalloproteins by almost 10 times, compared to analogous semi-classical models under certain conditions. Our model also extends the study of ET to temperatures that are too low for thermally activated stochastic hopping models to apply.
For a flat reduction potential profile in complex I, we predict an increased ET rate at moderate dephasing that is attributable to quantum coherence and its interplay with decoherence processes. For reduction potential profiles that vary markedly from site-to-site, our model converges with well-established semi-classical treatments of ET~\cite{Jortner:76}. 
We note that reduction potential measurements take place on far longer timescales than ET. Consequently,
the `transient' reduction profile relevant to ET may not include energetic contributions from slow relaxation processes. This could reconcile the requirement of quantum coherence for a flat reduction potential profile with the disorder in reduction potentials generally observed in biological molecules.
The quantum coherent enhancement of the ET rate in complex I is reminiscent, though not directly analogous to, recent studies of coherent exciton dynamics within photosynthetic complexes~\cite{Giorda:11, Chin:2010, Rebentrost:09, Asadian:2010}. Our model is general enough to be applied to other metalloproteins and can be adapted to investigate other non-biological ET processes also.

\section*{ACKNOWLEDGMENTS}
R.D. and J.G. wish to acknowledge N. Lo Gullo for helpful discussions.

\appendix

\section{Lang-Firsov transformation}

We show that  a more general Hamiltonian may be used as the starting point of our study than that of Eq.~\eqref{eq:initham}. We also note that it may serve as the starting point to investigate the ET dynamics of metalloproteins other than  NADH:ubiquinone oxidoreductase (complex I). The extended Holstein Hamiltonian~\cite{Holstein:59} can be partitioned into electronic, vibrational and interaction components ($H_\textrm{el}$, $H_\textrm{vib}$ and $H_\textrm{int}$ respectively) as
\begin{equation}
H=H_\textrm{el}+H_\textrm{vib}+H_\textrm{int}.
\label{eq:hampart}
\end{equation}
The electronic part of the Hamiltonian describes an N site tight-binding lattice with disorder in both the on-site energies $E_i$ and the nearest-neighbour tunneling amplitude $t_{j,j+1}$
\begin{equation}
H_\textrm{el}=\sum_{i}^N E_i c_i^\dagger c_i+h\sum_{j}^{N-1}  t_{j,j+1} \left( c^\dagger_j c_{j+1} +c^\dagger_{j+1} c_j \right).
\label{eq:hamelec}
\end{equation}
The vibrational component, $H_\textrm{vib}$, describes a set of harmonic oscillators associated with each of the N lattice sites. The $k_i^{th}$ mode at site $i$ has a characteristic frequency $\nu_{i,k_i}$ and corresponding bosonic creation and annihilation operators, $a^\dagger_{i,k_i}$ and $a_{i,k_i}$
\begin{equation}
H_\textrm{vib}=h\sum_{i}^N \sum_{k_i} \nu_{i,k_i}  a^\dagger_{i,k_i} a_{i,k_i}.
\label{eq:hamvibr}
\end{equation}
Finally, the interaction term $H_\textrm{int}$ accounts for the vibronic coupling
\begin{equation}
H_\textrm{int}=
h\sum_{i}^N\sum_{k_i} g_{i,k_i}  c^\dagger_{i} c_{i} \left(a_{i,k_i} + a^\dagger_{i,k_i} \right).
\label{eq:interham}
\end{equation}
Where $g_{i,k_i}$ is the coupling strength between the $k_i^{th}$ phonon mode and an electron at site $i$. The vibronic coupling thus describes the displacement of the phonon modes at a given site conditioned on the presence of an electron at that location.

Combining Eqs.~\eqref{eq:hampart} - \eqref{eq:interham} we obtain the full Holstein Hamiltonian
\begin{align}
H={} & H_\textrm{el}+H_\textrm{vib}+H_\textrm{int}.
\nonumber \\
= {} &\sum_{i}^N E_i c_i^\dagger c_i+h\sum_{j}^{N-1}  t_{j,j+1} \left( c^\dagger_j c_{j+1} +c^\dagger_{j+1} c_j \right)+h\sum_{i}^N\sum_{k_i} \nu_{i,k_i}  a^\dagger_{i,k_i} a_{i,k_i}
\nonumber \\
& +  h\sum_{i}^N\sum_{k_i} g_{i,k_i}  c^\dagger_{i} c_{i} \left(a_{i,k_i} + a^\dagger_{i,k_i} \right).
\label{eq:holham}
\end{align}

The Holstein Hamiltonian is often treated using a polaron approach, facilitated by the Lang-Firsov~\cite{Lang:63} (or `polaron') transformation using the unitary operator
\begin{align}
V&=e^S,
\nonumber \\
S&=\sum_i \sum_{k_i} \frac{g_{i,k_i}}{\nu_{i,k_i}}c^\dagger_i c_i \left( a^\dagger_{i,k_i} - a_{i,k_i}\right).
\end{align}
The transformed fermionic operators are calculated with use of the Hadamard lemma\footnote{For $A,B \in \mathbb{C}^{m \times m}$, the Hadamard lemma states: $e^A B e^{-A}=B+\left[A,B \right]+\frac{1}{2!}\left[ A,\left[ A,B \right] \right]+\frac{1}{3!}\left[A, \left[ A,\left[ A,B \right] \right] \right]+ \ldots$}
\begin{align}
\bar{c}_{j}= {} & V c_j V^\dagger,
\nonumber \\
= {} & c_j \textrm{exp}\left(-\sum_{k_j} \frac{g_{j,k_j}}{\nu_{j,k_j}} \left( a^\dagger_{j,k_j} - a_{j,k_j} \right) \right).
\label{eq:knockhole1}
\\
\rightarrow \bar{c}^\dagger_{j} = {} &c^\dagger_j \textrm{exp}\left(\sum_{k_j} \frac{g_{j,k_j}}{\nu_{j,k_j}} \left( a^\dagger_{j,k_j} - a_{j,k_j} \right) \right).
\label{eq:knockhole2}
\end{align}
Where Eq.~\eqref{eq:knockhole2} follows from hermitian conjugation of Eq.~\eqref{eq:knockhole1}.

Similarly, the transformed bosonic operators are
\begin{align}
\bar{a}_{j,k_j}={} & V a_{j,k_j} V^\dagger,
\nonumber \\
= {} & a_{j,k_j}-\frac{g_{j,k_j}}{\nu_{j,k_j}}c^\dagger_j c_j.
\label{eq:knockhole3}
\\ 
\rightarrow \bar{a}^\dagger_{j,k_j} = {} & a^\dagger_{j,k_j}-\frac{g_{j,k_j}}{\nu_{j,k_j}}c^\dagger_j c_j.
\label{eq:knockhole4}
\end{align}
With  Eq.~\eqref{eq:knockhole4} following from hermitian conjugation of Eq.~\eqref{eq:knockhole3}.
Combining Eqs.~\eqref{eq:knockhole1} - \eqref{eq:knockhole4}, the Lang-Firsov transformed Holstein Hamiltonian is found to be
\begin{align}
\bar{H} ={}&V H V^\dagger,
\nonumber \\
={}&\sum_i^N E^*_i c^\dagger_i c_i+h\sum_{j}^{N-1} t_{j,j+1} \left(c^\dagger_j c_{j+1} \bar{X}_{j}^\dagger \bar{X}_{j+1} +c^\dagger_{j+1} c_j \bar{X}_{j+1}^\dagger\bar{X}_{j} \right)
\nonumber \\
& +h \sum_i^N \sum_{k_i} \nu_{i,k_i} a_{i,k_i}^\dagger a_{i,k_i}.
\label{eq:transhamapp}
\end{align}
Where
\begin{align}
E_i^*&=E_i- \sum_{k_i}\frac{h g_{i,k_i}^2}{\nu_{i,k_i}},
\label{eq:renormenergy} \\
\bar{X}_{i} &= \textrm{exp}\left(- \sum_{k_i}\frac{g_{i,k_i}}{\nu_{i,k_i}}\left(a_{i,k_i}^\dagger-a_{i,k_i} \right)\right). 
\end{align}

\section{Diagonal processes}

We provide a derivation of the renormalised coherent tunneling amplitude arising from `diagonal processes' - in which the number of phonons remains unchanged within the chain upon a tunneling of an electron. The following derivation follows that of Ref.~\cite{Mahan:00} closely.
Beginning with Eq.~\eqref{eq:effham}
\begin{equation}
\bar{H}=\sum_i^N E^*_i c^\dagger_i c_i+h\sum_{j}^{N-1} t_{j,j+1} \left(c^\dagger_j c_{j+1} \bar{X}_{j}^\dagger \bar{X}_{j+1} +c^\dagger_{j+1} c_j \bar{X}_{j+1}^\dagger\bar{X}_{j} \right)+h \sum_i^N \nu_{i} a_{i}^\dagger a_{i}.
\nonumber
\end{equation}
Where, now
\begin{align}
E_i^*&=E_i-\frac{hg_{i}^2}{\nu_{i}},
\label{eq:newrenormenergy} \\
\bar{X}_{i} &= \textrm{exp}\left(- \frac{g_{i}}{\nu_{i}}\left(a_{i}^\dagger-a_{i} \right)\right). 
\label{eq:transdis}
\end{align}
Considering the case when $t_{j,j+1} \ll E_i, \nu_i$, the polaron tunneling term of Eq.~\eqref{eq:effham} may be treated as a small perturbation. To first order of the perturbation expansion we must evaluate terms of the form
\begin{equation}
\sum_{j}^{N-1}t_{j,j+1}\bra{\{n\}}\bra{i}  \left(c^\dagger_j c_{j+1} \bar{X}_{j}^\dagger \bar{X}_{j+1} +c^\dagger_{j+1} c_j \bar{X}_{j+1}^\dagger\bar{X}_{j} \right)\ket{\{m\}}\ket{i+1}
\label{eq:anobject}
\end{equation}
Where $\ket{\{n\}}$ ($\ket{\{m\}}$) is a multi-mode Fock state describing the phonon occupation numbers in the initial (final) state by the set of numbers $\{n\}=n_1,n_2...n_N$ ($\{m\}=m_1,m_2...m_N$), and $\ket{i}$ denotes the state of an electron located at site $i$.
The diagonal contribution $t^*_{i,i+1}$ to the object in Eq.~\eqref{eq:anobject} amounts to the evaluation of the following quantity
\begin{equation}
t^*_{i,i+1}=\sum_{j}^{N-1}\bra{\{n\}}\bra{i} t_{j,j+1} \left(c^\dagger_j c_{j+1} \bar{X}_{j}^\dagger \bar{X}_{j+1} +c^\dagger_{j+1} c_j \bar{X}_{j+1}^\dagger\bar{X}_{j} \right)\ket{\{n\}}\ket{i+1}.
\label{eq:sorenose}
\end{equation}
Factoring the fermionic and bosonic parts of Eq.~\eqref{eq:sorenose} gives
\begin{align}
t^*_{i,i+1}&=\sum_{j}^{N-1}t_{j,j+1}\left(\bra{i} c^\dagger_j c_{j+1} \ket{i+1} \bra{\{n\}} \bar{X}_{j}^\dagger \bar{X}_{j+1} \ket{\{n\}} +\bra{i}c^\dagger_{j+1} c_j \ket{i+1}\bra{\{n\}}\bar{X}_{j+1}^\dagger\bar{X}_{j} \ket{\{n\}} \right),
\nonumber \\
&=t_{i,i+1}  \bra{\left\{ n \right\}}  \textrm{exp}\left( \frac{g_{i}}{\nu_{i}}\left(a_{i}^\dagger-a_{i} \right)\right) \textrm{exp}\left( -\frac{g_{i+1}}{\nu_{i+1}}\left(a_{i+1}^\dagger-a_{i+1} \right)\right) \ket{\left\{ n \right\}}.
\label{eq:arandnam}
\end{align}
Where the second line follows from the orthonormality condition $\bra{j}\ket{i}=\delta_{i,j}$.
Applying the Baker-Campbell-Haussdorff-Zassenhaus formula\footnote{For $A,B \in \mathbb{C}^{m \times m}$, $r,t \in \mathbb{C}$, $\left[A,B \right]=r$, the BCHZ formula reduces to: $\textrm{exp}\left(t \left( A+B \right)\right)=\textrm{exp}\left(tA\right)\textrm{exp}\left(tB\right)\textrm{exp}\left(-\frac{t^2}{2!}r\right)$.} to Eq.~\eqref{eq:arandnam} gives
\begin{align}
t^*_{i,i+1}={}& t_{i,i+1}\textrm{exp}\left( -\frac{1}{2}\left(\frac{g^2_{i}}{\nu^2_{i}}+\frac{g^2_{i+1}}{\nu^2_{i+1}}  \right) \right)\bra{n_{i}}\textrm{exp}\left(\frac{g_i}{\nu_i} a_i^\dagger \right) \textrm{exp}\left(-\frac{g_i}{\nu_i} a_i \right)\ket{n_{i}}\bra{n_{i+1}} \textrm{exp}\left(-\frac{g_{i+1}}{\nu_{i+1}} a_{i+1}^\dagger \right) \textrm{exp}\left(\frac{g_{i+1}}{\nu_{i+1}} a_{i+1} \right)  \ket{n_{i+1}}
\label{eq:bigsigh}
\end{align}
Where, the $\ket{n_i}$ are single mode Fock states describing the vibrational state of the $i^{th}$ site. To proceed, the thermal average of the quantity in Eq.~\eqref{eq:bigsigh} is taken,
\begin{align}
t^*_{i,i+1}={}& t_{i,i+1}\textrm{exp}\left( -\frac{1}{2}\left(\frac{g^2_{i}}{\nu^2_{i}}+\frac{g^2_{i+1}}{\nu^2_{i+1}}  \right) \right) \frac{1}{\mathcal{Z}_{i}} \sum_{n_i=0}^\infty \textrm{exp}\left(-\frac{n_i h \nu_i}{k_BT}\right) \bra{n_{i}}\textrm{exp}\left(\frac{g_i}{\nu_i} a_i^\dagger \right) \textrm{exp}\left(-\frac{g_i}{\nu_i} a_i \right)\ket{n_{i}} \hdots
\nonumber \\
& \hdots \times \frac{1}{\mathcal{Z}_{i+1}}\sum_{n_{i+1}=0}^\infty \textrm{exp}\left(-\frac{n_{i+1} h \nu_{i+1}}{k_BT}\right) \bra{n_{i+1}} \textrm{exp}\left(-\frac{g_{i+1}}{\nu_{i+1}} a_{i+1}^\dagger \right) \textrm{exp}\left(\frac{g_{i+1}}{\nu_{i+1}} a_{i+1} \right)  \ket{n_{i+1}},
\label{eq:longway}
\end{align}
where $\mathcal{Z}_i=\left( 1-\textrm{exp}\left( -\frac{h \nu_i}{k_B T} \right) \right)^{-1}$ is the partition function for the $i^{th}$ vibrational mode.

Considering the object
\begin{equation}
\frac{1}{\mathcal{Z}}\sum_{n=0}^\infty \textrm{exp}\left(-\frac{n h \nu}{k_BT}\right) \bra{n} \textrm{exp}\left(\xi a^\dagger \right) \textrm{exp}\left(-\xi a \right)  \ket{n},
\label{eq:billyw}
\end{equation}
It can be shown (see Ref. \cite{Mahan:00} for further details) that
\begin{equation}
\frac{1}{\mathcal{Z}}\sum_{n=0}^\infty \textrm{exp}\left(-\frac{n h \nu}{k_BT}\right) \bra{n} \textrm{exp}\left(\xi a^\dagger \right) \textrm{exp}\left(-\xi a \right)  \ket{n}=\textrm{exp}\left( -\xi^2 \langle n\rangle \right),
\end{equation}
where $\langle n(T) \rangle=\left(\textrm{exp}\left(\frac{h \nu}{kT}\right)-1\right)^{-1}$ is the Planck average of the number of phonons.
Substituting this result into Eq.~\eqref{eq:longway}, finally gives the thermally averaged rate of diagonal transitions
\begin{align}
t^*_{i,i+1}= t_{i,i+1}\textrm{exp}\left( -\frac{g^2_{i}}{\nu^2_{i}}\left(\frac{1}{2}+\langle n_i(T) \rangle\right)+\frac{g^2_{i+1}}{\nu^2_{i+1}}\left(\frac{1}{2}+\langle n_{i+1}(T)\rangle  \right) \right)
\end{align}
We see that the vibronic coupling manifests itself as an exponential suppression of the polaronic tunneling amplitude $t^*_{j,j+1}$ relative to its `bare' electronic counterpart $t_{j,j+1}$. This is attributable to the effective mass acquired by the electron due to its coupling to phonons in the polaron picture. Polaron tunneling is also exponentially sensitive to the number of phonons excited in the relevant clusters; a quantity that increases as a function of temperature, leading to further suppression of the tunneling amplitude $t^*_{j,j+1}$.

\section{Born-Markov master equation}

Interaction of a quantum system with an environment introduces non-unitary processes to its dynamics. These manifest themselves in the form of local dephasing processes; that lead to the `decay' of superposition states to statistical mixtures of those states, and dissipation; the irreversible loss of a particle from the system. 
The description of these `incoherent' processes is aided by adopting the density matrix formalism\footnote{The reduced state of the system refers to the density matrix obtained after a partial trace is performed over the environmental degrees of freedom for the joint denisty matrix describing the state of the system and environment together (see for example \cite{pet:06}). In the case of complex I the state of the system may be written as $\ket{\psi(t)}=\sum_{j=0}^7 b_j(t) \ket{j}$, where the $b_j(t)$ are time dependent amplitudes. The reduced density matrix describing the system is then $\rho(t)=\ket{\psi(t)}\bra{\psi(t)}$.}. This is convenient when describing dissipative processes, as the diagonal elements of the density matrix can be identified as the `population' of the sites, while the off diagonal elements describe the coherence between different basis states.

The origin of local dephasing within our model can be associated with the outer sphere reorganisation energy $\lambda_\textrm{out}$ describing the interaction of the polaron with its surrounding protein scaffold and solvent. We argue that environmental vibrational modes interact only weakly with the polaron due to the screening of the electron charge by the distorted cluster ions, allowing us to invoke the Born approximation. The additional assumption of a Markovian environment allows the dephasing process to be written in the form of a Lindblad term
\begin{equation}
\mathcal{L}_\textrm{deph}\left(\rho(t)\right)=\sum_{i=1}^N \gamma \left( -\left\{ c^\dagger_i c_i ,\rho(t)\right\} +2c^\dagger_i c_i\rho(t) c^\dagger_i c_i \right),
\end{equation}
Where $\gamma=\gamma_0\left(1+k_B T/h \nu \right)$ is the rate of dephasing at all sites. In lieu of a detailed knowledge of the density of modes for the environment, we treat the projected magnitude of quantum dephasing at zero temperature, $\gamma_0$\ as a free parameter and scan the range $\gamma_0=0 - 1$ GHz to investigate the robustness of the polaron transport under dephasing.

The ultimate goal of ET in complex I is the reduction of a quinone molecule located at the end of the Fe-S chain. Once the quinone is reduced it dissociates from the enzyme. In this way the quinone acts as a sink and introduces a further non-reversible process to the dynamics of the system.
Dissipation of the polaron from the $N^{th}$ site at a rate $R$ is mediated by another Lindblad term
\begin{equation}
\mathcal{L}_\textrm{sink}\left(\rho(t) \right)= R \left( -\left\{ c^\dagger_7 c_7 ,\rho(t)\right \} +2c_7\rho(t) c^\dagger_7  \right).
\end{equation}
Dissipation from all other sites can be accounted for in a similar way, however the timescale for this process is taken to be much longer than the transport time of the electron, and as such is irrelevant to our model.

Incoherent tunneling enters our model via the Marcus-Jortner rate equation (see Eq.~\eqref{eq:marcjort}). This process accounts for transitions of the electron between two sites in which its phase coherence is not preserved, due to inelastic scattering with phonons. For sites $j$ and $j+1$, the Marcus-Jortner equation predicts a forwards rate $\alpha_{j \rightarrow j+1}$ and, by enforcing detailed balance, a backwards rate $\alpha_{j+1 \rightarrow j}$ of electron transfer. The Liouvillian term describing incoherent tunneling is thus
\begin{align}
\mathcal{L}_\textrm{tunn}\left(\rho(t)\right)&=\sum_{j=1}^{N-1} \alpha_{j \rightarrow j+1} \left( -\left\{  c^\dagger_j c_{j+1}c^\dagger_{j+1} c_j\rho(t) ,\rho(t)\right\} +2c^\dagger_{j+1} c_j\rho(t) c^\dagger_j c_{j+1} \right)
\nonumber \\
&+\alpha_{j+1 \rightarrow j} \left( -\left\{  c^\dagger_{j+1} c_{j}c^\dagger_{j} c_{j+1}\rho(t) ,\rho(t)\right\} +2c^\dagger_{j} c_{j+1}\rho(t) c^\dagger_{j+1} c_{j} \right).
\end{align}

The full evolution of the density matrix is then described by the Born-Markov master equation
\begin{equation}
\frac{d \rho(t)}{d t}=-\frac{i}{\hbar}\left[ H_\textrm{eff}, \rho(t) \right]+\mathcal{L}_\textrm{deph}(\rho)+\mathcal{L}_\textrm{sink}(\rho)+\mathcal{L}_\textrm{tunn}(\rho).
\label{eq:masteq}
\end{equation}

Where $H_\textrm{eff}$ is the Hamiltonian given in Eq.~\eqref{eq:transham}.

\section{Additional results}

\begin{figure}[tb]
\includegraphics[scale=0.6]{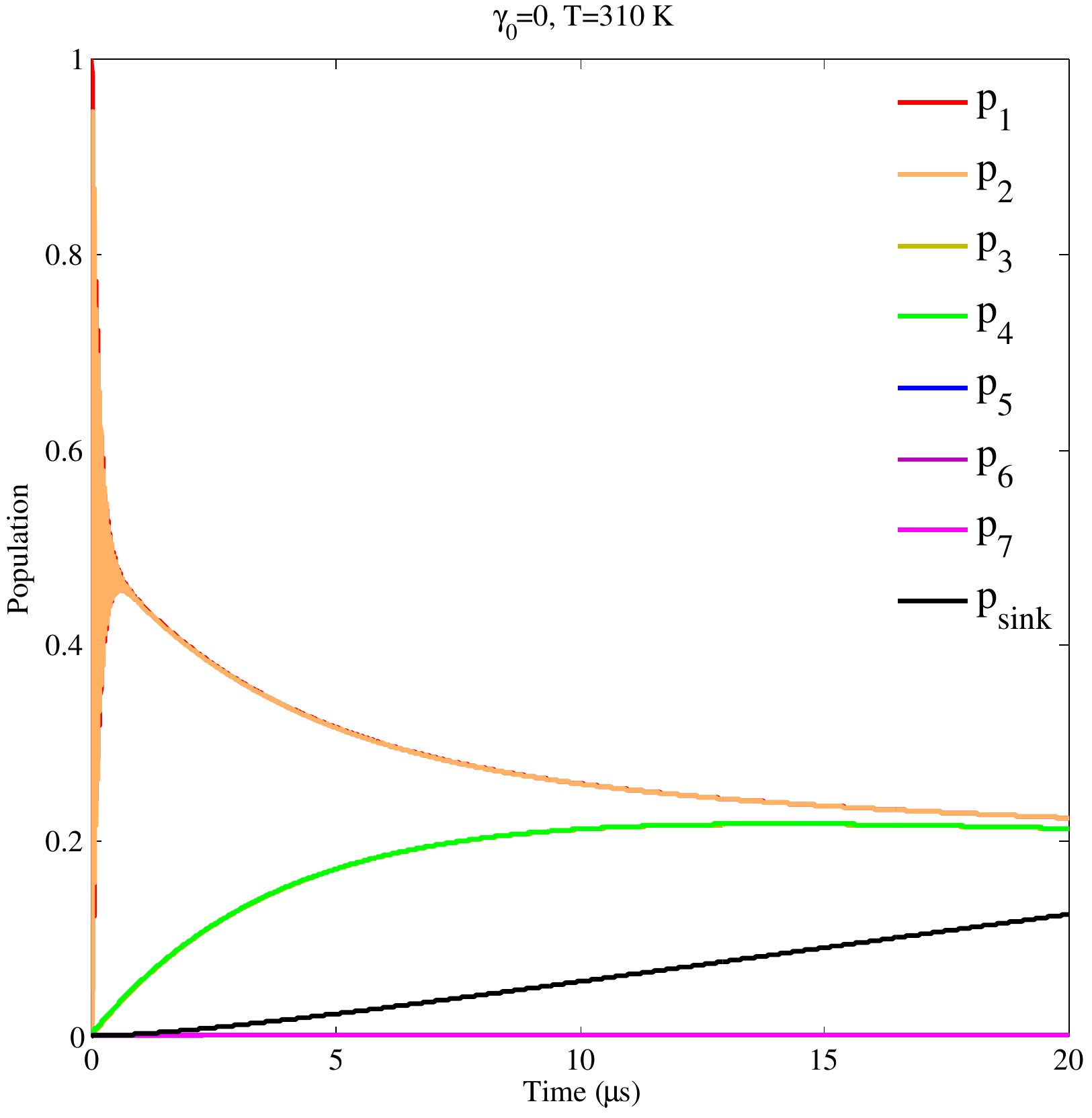}
\caption{The populations of the seven sites of complex I, and the probability of quinone reduction as a function of time with zero dephasing at $T= 310$ K. Coherent oscillations between sites 1 and 2 persist for several microseconds before the onset of qualitatively diffusion-like behaviour. The probability of quinone reduction does not exceed 0.25 after $40 \mu$s.}
\label{fig:supfig1}
\end{figure}

\begin{figure}
\centering
\mbox{\subfigure{\includegraphics[scale=0.5]{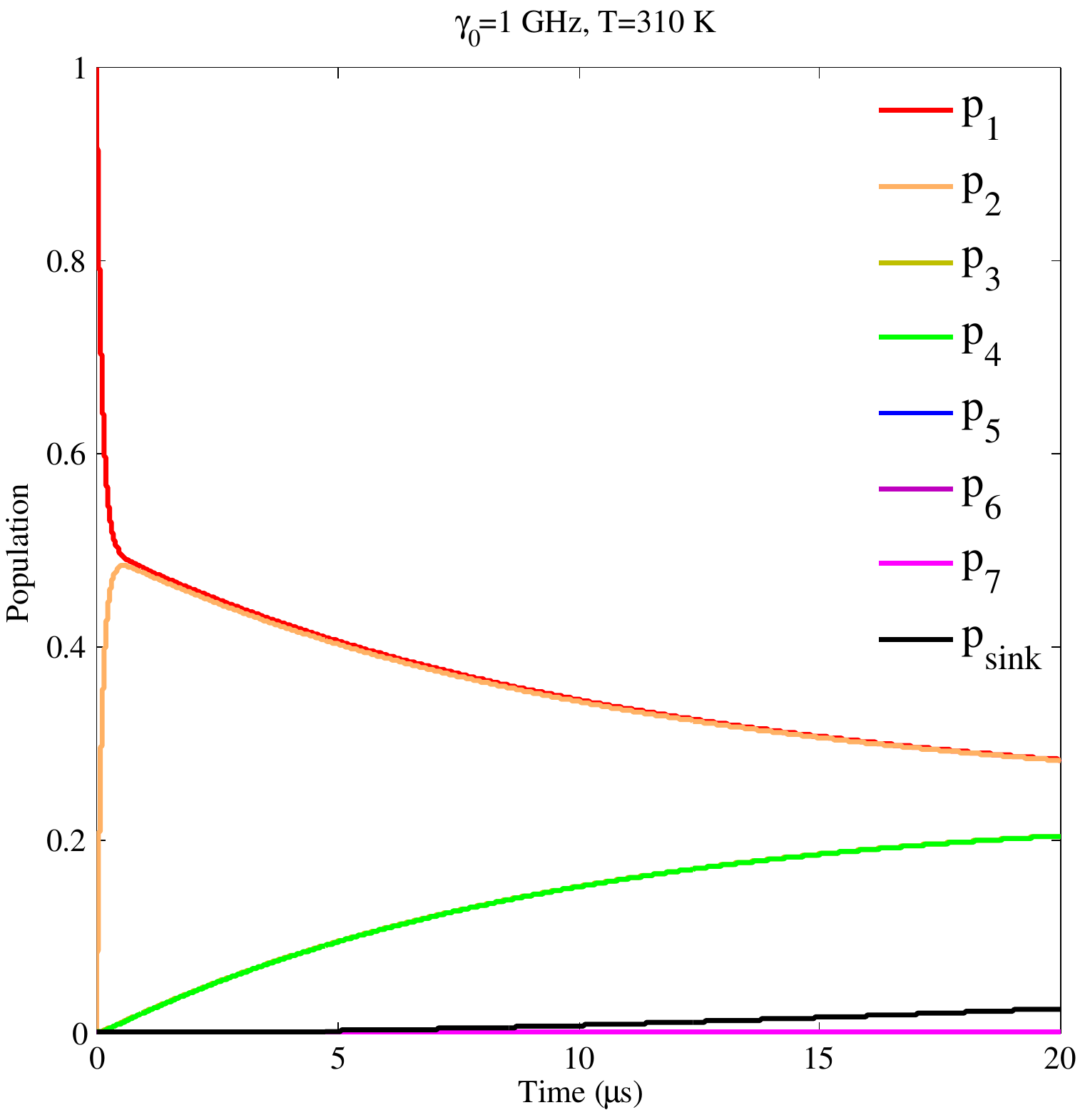}}\quad
\subfigure{\includegraphics[scale=0.5]{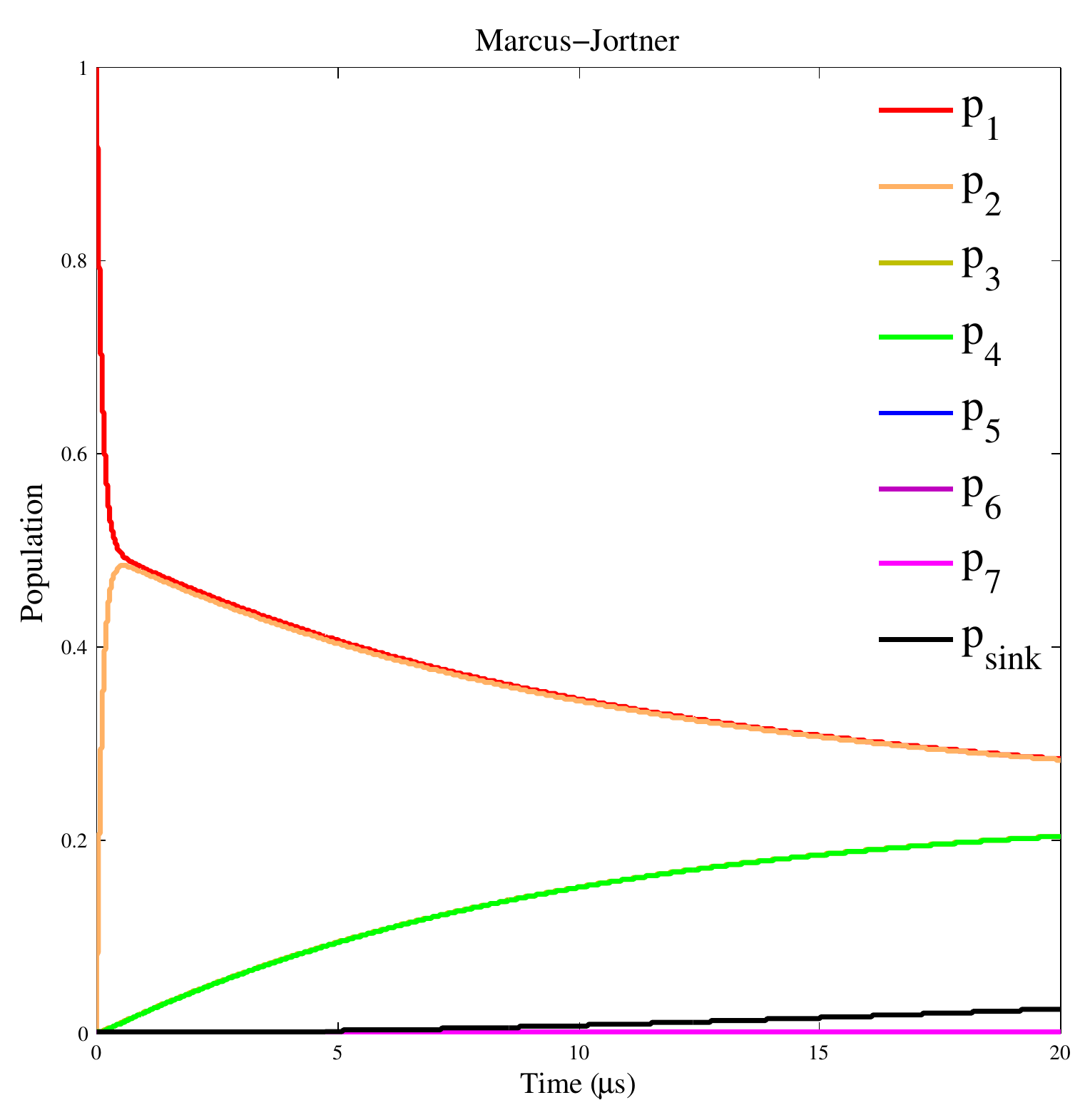} }}
\caption{{\bf Left:} The populations of the seven sites of complex I, and the probability of quinone reduction as a function of time under the influence of high dephasing ($\gamma_0>1$ GHz) at $T= 310$ K. Quantum coherence is not observed on a timescale relevant to ET and the typical Marcus-Jortner diffusive transport ({\bf right}) is recovered exactly.  The probability of quinone reduction does not exceed 0.1 after $40 \mu$s}
\label{fig:supfig2}
\end{figure}

Fig. \ref{fig:fig2} and Figs.~\ref{fig:supfig1} - \ref{fig:supfig2} show the effect of dephasing upon ET within complex I. For a low rate of dephasing ($\gamma_0<0.001$ GHz) at physiological temperature (see Fig.~\ref{fig:supfig1}), the presence of coherent oscillations between the populations of adjacent states is observable on microsecond timescales. These oscillations are the hallmark of quantum transport. The amplitude of the oscillations is damped primarily by the onset of incoherent tunneling and ET qualitatively approaches diffusive behavior on the longer timescale of quinone reduction.
For high rates of dephasing ($\gamma_0 >1$ GHz ) and $T=310$ K, quantum coherence does not persist on a timescale relevant to ET and the dynamics are purely diffusive. Fig.~\ref{fig:supfig2} shows that in the limit of large dephasing, the dynamics quantitatively approach the behaviour expected using Marcus-Jortner rates within a classical master equation. 

ET at intermediate rates of dephasing ($\gamma_0 \sim 0.001$ GHz) and $T = 310$ K exhibit coherent oscillations that persist for approximately 100 ns (see Fig.~\ref{fig:fig2}), approximately $100-1000$ times shorter than the typical timescale of quinone reduction. There are two noteworthy aspects of the ET in this regime: $i)$ The rate of quinone reduction is significantly enhanced compared to the cases of low (more `quantum') and high  (more `classical') dephasing. $ii)$ Quantum coherence plays a significant role in ET despite persisting for only a fraction of the typical timescale of quinone reduction ($\sim10-100 \mu$s). The enhancement of transport rate in the presence of intermediate dephasing arises from the interplay of quantum coherent and classical transport. This behavior is reminiscent, though not directly analogous to, the notion of `dephasing assisted transport'. In our model, the interplay of classical and quantum transport is richer due to the presence of incoherent hopping alongside pure dephasing.



\end{document}